\documentclass[12pt]{article}
\usepackage{epsfig}

\def\beq{\begin{equation}}
\def\eeq#1{\label{#1}\end{equation}}
\def\eeqn{\end{equation}}

\def\beqa{\begin{eqnarray}}
\def\eeqa#1{\label{#1}\end{eqnarray}}
\def\eeqan{\end{eqnarray}}

\let\bar=\overbar

\def\etal{{\it et al.}}

\def\Dslash{\not{\hbox{\kern-4pt $D$}}}
\def\dslash{\not{\hbox{\kern-2pt $\del$}}}

\def\msb{{\bar{\ssstyle M \kern -1pt S}}}

\def\Title#1{\begin{center} {\Large {\bf #1} } \end{center}}

\begin{document}

\Title{The BTeV experiment at the Tevatron collider}

\bigskip\bigskip


\begin{raggedright}  

{\it Jianchun Wang\index{Wang, J.C.} {\rm representing the BTeV Collaboration}\\
Department of Physics\\
Syracuse University\\
Syracuse, NY 13244}
\bigskip\bigskip
\end{raggedright}


\section{Physics Motivation}

The BTeV experiment is designed to study beauty and charm physics
at the Fermilab Tevatron collider.
Our goals are to make an exhaustive search for physics beyond the Standard Model (SM)
and make precise measurements of the SM parameters.
The important measurements to make involve CP violation, mixing, and rare decays
of hadrons containing {\it b} or {\it c} quarks.

The CP violation in the SM originates from quark mixing with complex terms in
the CKM matrix.
The unitarity of the CKM matrix allows us to construct 6 triangles.
The most commonly used triangle arises from the orthogonality
of the {\it d} and {\it b} columns:
$\ V_{ud}^*V_{ub} + V_{cd}^*V_{cb} + V_{td}^*V_{tb} = 0,$
which defines the the CKM phases $\alpha, \beta$ and $\gamma$ with the constraint that
$\alpha + \beta + \gamma = \pi$.
Other independent angles are $\chi$ and $\chi^\prime$ where:
$$ \chi=arg\left(-{V^*_{cs}V_{cb}\over V^*_{ts}V_{tb}}\right), \ \ \ 
   \chi'=arg\left(-{{V^*_{ud}V_{us}}\over {V^*_{cd}V_{cs}}}\right). $$
While $\alpha, \beta$ and $\gamma$ maybe relatively large, the 
angle $\chi$ is small, and $\chi^\prime$ is even smaller.
One goal of the BTeV experiment is to measure the CKM phases:
$\alpha, \beta, \gamma$ and $\chi$.

We usually measure a trigonometric function of the angles.
For example, the decay $B^\circ \rightarrow J/\psi K_s$ measures sin(2$\beta$).
There is a 4-fold ambiguity generated by conversion from sin(2$\beta$) to $\beta$.
The ambiguity can be reduced by finding cos(2$\beta$).

Inconsistencies in determinations of CKM parameters using different physics
processes may also reveal new physics. 
One good candidate is the decay mode $B^\circ \rightarrow \phi K_s$.
The CP asymmetry in the SM for this mode is the same as for
$B^\circ \rightarrow J/\psi K_s$.
In models containing new physics both the mixing amplitude and the
decay amplitude can be modified by new phases.  However, the $J/\psi K_s$
decay being tree level does not usually pick up any new phase, while the
loop level $\phi K_s$ decay is likely to have a new phase. Then the CP
asymmetry in
$B^o\rightarrow J/\psi K_s$ becomes proportional to $\sin(2\beta +
\theta_D)$, while in $B^o\rightarrow\phi K_s$ we have $\sin(2\beta +
\theta_D +\theta_A)$.
Thus a measurement of the difference in CP asymmetries between these two
modes would definitively demonstrate new physics and measure the decay
phase $\theta_A$ of the new physics~\cite{Nir:1999mg}.

The angle $\chi$ can be extracted by measuring the time dependent CP violation
asymmetry using CP eigenstates in $B_s$ decay modes.
Silva and Wolfenstein ~\cite{Silva:1996ih, Aleksan:1994if} show that the $\chi$ measurement
is a critical check to the SM by seeing if
$$\sin\chi = \left|{V_{us}\over 
V_{ud}}\right|^2{{\sin\beta~\sin\gamma}\over{\sin(\beta+\gamma)}}\ \ .$$
The BTeV can measure $\chi$ in the decay modes $B_s \rightarrow J/\psi \eta^{(\prime)}$,
where $\eta \rightarrow \gamma\gamma$ and $\eta^\prime\rightarrow \rho^\circ\gamma$,
or $\pi^+\pi^-\eta$.
Since $\chi$ is very small ($\approx 0.03$), we need several years data collection
to have a reasonable precise measurement.

\section{The One-arm Spectrometer}

\begin{figure}[htbp]
\centerline{\epsfig{figure=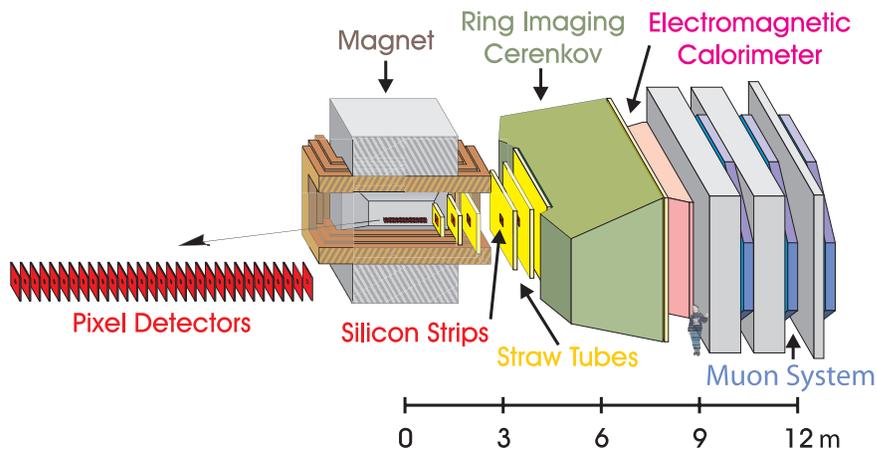,height=2.3in}}
\caption{\label{btev_det} Layout of the BTeV spectrometer.}
\end{figure}

The BTeV detector is an one-arm spectrometer covers the angular region between 
10-300 mrad with respect to the beam as shown in Fig.~\ref{btev_det}.
It fully exploits two advantages of the ``forward'' direction: the correlation in
the direction of the $b\bar{b}$ pair produced, and the boost that allows an easier
identification of detached vertices.
This provides efficient flavor tagging and sensitivity to a great variety of heavy
flavor decays.

The spectrometer consists of:
a planar precision vertex detector made from silicon pixels,
a forward tracking system comprised of silicon strips near the beam
and straw tube chambers at larger radius,
a Ring-Imaging Cherenkov Detector (RICH), an $\rm PbWO_4$ electromagnetic calorimeter,
a muon system, and a deadtimeless trigger and DAQ system.
Here we will give a brief description on some key sub-detectors.
More information can be found elsewhere~\cite{Wang:fk}.

The pixel vertex detector locates inside an 1.6 Tesla dipole magnetic field~\cite{Kwan:hi}.
It provides accurate vertex information for offline analysis, 
and delivers very clean, precision space points to the vertex trigger.
The pixel vertex detector consists of 30 stations of doublets along the beam direction,
with pixel size $\rm 50\times 400 \mu m^2$.
The pixel geometry is chosen to provide excellent signal-to-noise, spatial resolution, 
high speed and low occupancy.
A beam test of prototype pixel detectors had been carried on in 1999-2000 at Fermilab.
The resolution is excellent, better than the 9 $\rm \mu m$ requirement~\cite{Appel:2001xd}.
A simulation package describes the detector performance quite well~\cite{Artuso:2000nh}.

The BTeV RICH detector is designed to separate $\pi/K/p$ in a momentum range 
of 3 to 70 GeV~\cite{Skwarnicki:hf}.
It is essential to CP violation studies, providing separation of specific final states,
such as $K^+\pi^-$ from $\pi^+\pi^-$, and flavor tagging.
The RICH detector is also a fine supplement to the electromagnetic calorimeter and
the muon detector in lepton identification. 
It increases significantly the reconstruction efficiency in interesting modes like
$B^\circ \rightarrow J/\psi K_s$.
We use freon ($\rm C_4 F_{10}$) gaseous radiator to generate Cherenkov light
in the optical frequency range.
The light is focused by mirrors onto Hybrid Photo-Diode (HPD) tubes at upstream.
To separate kaons from protons below the threshold of gaseous radiator, 
an liquid radiator ($\rm C_5 F_{10}$) is used and the light is focused onto the side
of the vessel and detected by PMT array.

BTeV uses radiation hard lead tungsten scintillation crystals ($\rm PbWO_4$)
to detect photons and electrons~\cite{Menary:dp}.
The crystals are 220 mm long (25 $\rm X_0$) and have very small transverse
cross-section ($\rm \approx 28\times 28 mm^2$), providing excellent segmentation.
The light is collected by PMT.
Sample crystals were tested in a beam at Protvino, Russia. 
The energy and position resolutions were demonstrated to be excellent.

With a nominal luminosity of $\rm 2 \times 10^{32} cm^{-2}s^{-1}$, the Tevatron
delivers $\rm 2 \times 10^{11}$ $b$-hadron per year.
It is an ideal place for study of $b$-physics.
It is also a severe challenge to the data acquisition (DAQ) system:
Only a small portion of inelastic events are from heavy quarks,
and the interaction rate is very high at the bunching spacing of 132 ns.

The BTeV trigger system mainly relies on the sophisticated detached vertex trigger,
using lifetime to distinguish ``B'' events from others~\cite{Lebrun:gf}.
It uses hits provided by the pixel vertex detector.
The parallel pipelines are implemented at Level 1, thus it is deadtimeless.
At Level 2 and Level 3, there is much more time to have refined reconstruction.
With the robust system, the total event rate is reduced from 7.6 MHz to 2~-~4 kHz.
The trigger efficiency for a typical beauty decays is greater than 50\%.

\section{Physics Reach}

We used GEANT-based Monte Carlo simulation to calculate the physics reach.
A full pattern recognition was done at trigger level.
Although there are fewer resources at this level than in offline analysis,
the algorithm was demonstrated to have high efficiency and very few false tracks.
For ``offline'' analysis, pattern recognition is not included since it has
very little impact due to the excellent segmentation of the pixel detector.
The charged tracks are fitted with Kalman filter.
Realistic shower reconstruction, particle identification programs are also used.
The results on CKM parameters for $10^7$s running at nominal luminosity are
shown in Table~\ref{tab:CKM}.
The accuracies in all cases are quite promising.
\begin{table}
\begin{tabular}{lrrcc}
\hline
Decay modes & \# of Events & S/B & Parameter & Error or (Value) \\ \hline
$B^{\circ}\to\pi^+\pi^-$   &   14,600   &   3   &   Asymmetry    &   0.030\\
$B_s\to D_s^+ K^-$         &    7,500   &   7   &   $\gamma$     &  8$^{\circ}$\\
$B^{\circ}\to J/\psi K_s$  &  168,000   &  10   & $\sin(2\beta)$ & 0.017\\
$B_s\to D_s^+\pi^-$        &   59,000   &   3   &   $x_s$        & (75) \\
\hline
$B^-\to\overline{D}^{\circ}(K^+\pi^-)K^-$ & 170 & 1 &            &      \\
$B^- \to {D}^{\circ}(K^+K^-)K^-$ & 1,000 & $>10$&   $\gamma$     & 13$^{\circ}$\\
\hline
$B^-\to K_s\pi^-$          &    4,600   &   1   &                & $<4^{\circ}$ + \\
$B^{\circ}\to K^+\pi^-$    &   62,100   &  20   & $\gamma$       & theory errors \\
\hline
$B^\circ\to\rho^+\pi^- $   &    5,400   & 4.1   &                &  \\
$B^\circ\to\rho^\circ\pi^\circ$&  780   & 0.3   &    $\alpha$    & $\sim 4^{\circ}$\\
\hline
$B_s\to J/\psi \eta$       &    2,800   &  15   &                &\\
$B_s\to J/\psi \eta'$      &    9,800   &  30   &$\sin (2\chi)$  & 0.024 \\
\hline
\end{tabular}
\caption{BTeV physics reach in CKM parameters for $10^7$s data}
\label{tab:CKM}
\end{table}

The BTeV data samples will be large enough to test new physics.
In Table~\ref{tab:newphys} we list some of the event samples relevant to new physics 
studies with one ``snow mass'' year data collection. 
Also shown in comparison are samples from $e^+e^-\ B$ factories that has 500 $\rm fb^{-1}$ 
accumulated data, approximately the expected amount when BTeV starts.
The BTeV will easily surpass the $e^+e^-\ B$ factories and also have accesses to
the important CP violation measurements that need to be made in $B_s$ modes.

\begin{table}
\begin{tabular}{lrrrrrr}
\hline
\vspace{1mm}
\bf{Decay modes} & \multicolumn{3}{c}{\bf BTeV ($10^7$ s)}& 
\multicolumn{3}{c}{\bf $B$-factories (500 fb$^{-1}$)}\\
                          &  Yield  & Tagged & S/B   & Yield& Tagged& S/B \\\hline
$B_s\to J/\psi\eta^{(')}$ & 12,650  & 1,645  & $>15$ &   0 &     &  \\
$B^-\to \phi K^-$         &  6,325  & 6,325  & $>10$ & 700 & 700 & 4\\
$B^o\to \phi K_s$         &  1,150  &   115  & 5.2   & 250 &  75 & 4\\
$B^o\to K^{*o}\mu^+\mu^-$ &  2,530  & 2,530  & 11    & $\sim$50 &$\sim$50 & 3\\
$B_s\to \mu^+\mu^-$       &      6  &   0.7  & $>15$ &   0 &     &  \\
$B^o\to \mu^+\mu^-$       &      1  &   0.1  & $>10$ &   0 &     &  \\
$D^{*+}\to\pi^+D^o$, $D^o\to K^-\pi^+$ & $\sim 10^8$& $\sim 10^8$& large&
$8\times 10^5$&$8\times 10^5$ & large\\
\hline
\end{tabular}
\caption{Reconstructed events in new physics modes for BTeV and
$e^+e^-$ $B$-factories}
\label{tab:newphys}
\end{table}

The LHCb experiment will be the main competitor to the BTeV experiment.
Compare to LHCb, BTeV has more robust trigger system that use detached 
vertex trigger at the first level. 
In the ``golden'' mode $B_s \to D_s K^-$ for $\gamma$ measurement, although
LHCb has initial advantages in $b$ cross-section, the yield and signal to
background ratio are compatible between the two experiments.
The BTeV possesses much better electromagnetic calorimeter.
And thus it is superior in $\alpha$ measurement that relies on
$B^\circ\to\rho^+\pi^-, \rho^\circ\pi^\circ$ decay modes.

\section{Conclusions}

Many R\&D activities are going on. 
Test beam runs at Fermilab with the Pixel and Muon systems 
and at Protvino with the EM calorimeter have been very successful and will continue.
System tests on RICH, straws and silicon will be carried out soon.
Progress has been made on the trigger and DAQ. 
We have received IT funding from the NSF to develop fault-tolerant,
fault-adaptive software to control the trigger system in real time.
This is named ``The Real Time Embedded Systems'' (RTES) project. 

I quote the PAC recommendation here as conclusion:
``... BTeV has designed and prototyped an ambitious trigger that will use 
$B$ decay displaced vertices
as its primary criterion. This capability, together with BTeV's excellent 
electromagnetic calorimetry and particle ID and enormous yields, will allow
this experiment to study a broad array of $B$ and $B_s$ decays. BTeV has a 
broader physics reach than LHCb and should provide definitive measurements 
of CKM parameters and the most sensitive tests for new physics in the flavor sector.''

\end{document}